\documentclass[supp]{new_tlp}

\usepackage{aopmath}
\usepackage{graphicx}
\usepackage{subfigure}
\usepackage{url}
\usepackage{supp}

\graphicspath{{./figures/}}

\begin{document}
\bibliographystyle{acmtrans}

\long\def\comment#1{}

\title{Towards an Efficient Prolog System by Code Introspection}

\author[George S. Oliveira and Anderson F. da Silva]
{George S. Oliveira and Anderson F. da Silva\\
Departament of Informatics\\
State University of Maring\'a, Brazil\\
\email{geo.soliveira@gmail.com, anderson@din.uem.br}
}

\pagerange{\pageref{firstpage}--\pageref{lastpage}}
\volume{\textbf{10} (3):}
\jdate{March 2002}
\setcounter{page}{1}
\pubyear{2002}

\pubauthor{Zhang}
\jurl{xxxxxx}
\pubdate{22 June 2013}

\maketitle

\label{firstpage}

\begin{abstract}
To appear in Theory and Practice of Logic Programming (TPLP).
Several Prolog interpreters are based on the Warren Abstract Machine (WAM), an elegant model to compile Prolog programs. In order to improve the performance several strategies have been proposed, such as: optimize the selection of clauses, specialize the unification, global analysis, native code generation and tabling. This paper proposes a different strategy to implement an efficient Prolog System, the creation of specialized emulators on the fly. The proposed strategy was implemented and evaluated at YAP Prolog System, and the experimental evaluation showed interesting results.
\end{abstract}


\section{Introduction}
\label{sec.intro}

Several programming languages provide introspection, a way for programs to reason about their own internal structure. Besides in programming languages, introspection has been employed in dynamic translation systems \cite{INTROSPEC_DYNTRANSLATION}, security and reliability \cite{INTROSPEC_RELIABILITY} and deadlock detection \cite{INTROSPEC_DEADLOCK}.

Prolog provides a significant form of introspection by the \emph{clause} predicate. It allows the user to find metadata related to a goal in the database. Besides, using this predicate it is possible to create metacircular interpreters and write evaluators that use nonstandard search orders. 

The use of metadata provided by introspection should not be limited to outside tools. So that, this paper proposes the use of introspection (and metadata) to build an efficient Prolog System. Introspection can allow the system to reason about its internal structure and self specialize on the fly. This strategy was implemented and evaluated on YAP \cite{YAPPROLOG}, and the results indicated that is possible to reduce the emulator's overhead.

This paper is organized as follows. Section \ref{sec.related} presents related work. Section \ref{sec.architecture} describes the proposed strategy to construct an efficient Prolog System. Section \ref{sec.results} presents experimental results and Section \ref{sec.conclusion} ends with the concluding remarks. 

\section{Related Works}
\label{sec.related}

Several researches have been conducted with the purpose of improving the WAM performance. These include researches about optimization of selection of clauses \cite{DEMANDINDEXING}, specialization of unification \cite{COMPILCOMPTERMS}, global analysis \cite{CIAO}, generation of efficient code \cite{PARMA} and tabling \cite{YAPPROLOG}, which are, in part, already employed in current Prolog system. 

These researches, however, have their drawbacks: some of them require a refinement of the WAM (or variants) instruction set, or are very difficult to implement, or require the use of another technique to be truly effective, or are effective only for programs that manipulate large amount of data, in particular those for deductive database.

The purpose of this paper is to add a new strategy in order to improve the Prolog system performance, the use of introspection to reason about the engine's internal structure and based on it to generate on the fly specialized emulators.

\section{A Prolog System by Code Introspection}
\label{sec.architecture}

Prolog emulators have provided an attractive system with a simple and elegant language. Such systems have been used on several applications such as theorem proving \cite{TEOREMPROVING}, answer set programming \cite{TLP:9156582}, heap solver \cite{journals/tplp/AlbertBGRS13}, and others. Therefore, it is important to provide a system that can be specialized to the current program in order to minimize the intrinsic overhead of emulation. It is the purpose of this paper, to describe a Prolog System that uses introspection in order to create specialized emulators on the fly.

The Prolog language, although not be fully introspective, provides a degree of introspection. In Prolog, 
it is possible to execute the following example:
 
\begin{verbatim}

degree(rio, 36)
degree(maringa, 34)
degree(curitiba, 24)

hotsummer(X) :- degree(X, Y), Y > 30.

?- clause(hotsummer(X),B)
B = degree(X, Y), Y > 30

\end{verbatim}

This example demonstrated how Prolog provides a custom database perusal. A similar functionality can be used in the engine level. Through introspection, the engine can peruse its own metadata and use them to create mutable emulators.

Reasoning about its internal structure makes feasible not only return the control flow graph of a specific instruction, but also the basic blocks, the number of basic blocks, the size of each basic block, among other metadata. With this metadata, the system can build a new emulator, more precisely, a specialized emulator (hereafter called S.emulator).

It is important to note that such feature raises some questions, such as: 

\begin{itemize}
\item How can the system provide introspection? 
\item How does the generation of a specialized emulator occur?
\item What is the mutability and how does the system manage it?
\end{itemize}

\subsection{Providing Introspection}

In a context in which the emulator source code is available, it is possible to feed the system with metadata describing each instruction, besides the structure of the system. On the other hand, if the source code is not available it is possible to use some strategy that comes from dynamic binary translation \cite{VIRTUALMACHINES}, to perform this task. Therefore, the ultimate goal is to capture during execution time this metadata, specialize them (if necessary), and thus generating a specialized emulator that be able to reduce the overhead of emulation.

A simple and effective strategy is to create specialized emulators in a conservative way to assure it will never reach an unsafe state. Such task has the potential to increase the interpretation's overhead. Therefore, it is not desirable that such task be enabled throughout the whole execution of the program. Ideally, it should only be enabled when the program reaches a state in which type of its structures is consistent. As a result, it is possible to generate a sYAAM emulator and ensure that an unsafe condition will not be met.

Adding metadata to the system implies on the addition of new features such as capacity of perusing and marking such metadata. At this point, a critical question arises: how does the system begin to reason about its metadata?

In the context of just-in-time compilation, the compilation system is achieved when a certain piece of code is invoked frequently, which is determined with the use of counters. In this case, a piece of code must be compiled when its amount of invocations exceeds a threshold.

Using the same approach, the proposed strategy instruments the predicates with counters and uses them  to initialize the task of perusing metadata. Therefore, when a counter reaches a certain threshold, the system begins to reason about its internal structure and marks the basic blocks executed by the current emulator instruction. It is important to note that the use of counters incurs in a minimal overhead because only one increment is executed in the head of the current clause (or fact). However, the markup task has an overhead that can negatively impact the system. Due to the task of perusing the metadata of the current instruction in order to find the current basic block and then mark it. Therefore, the proposed strategy provides a mechanism to enable and disable the markup task.

Disabling the markup task occurs when the system is running a S.emulator. If the system requires a new specialization, which will cause a return to the default emulator, the system enables the markup task and begisns the process of creating a new S.emulator. It is important to note that the system handles several emulators, however only the default emulator is the one be able to handle all the instructions and exceptions. On the other hand, a S.emulator does not have this capacity in order to be a specialized emulator.

\subsection{Creating a S.emulator on the Fly}

At the implementation level, every emulator instruction contains basic blocks that will always be executed and others that will only be executed from conditional branches. Eliminating conditionals branches and build a S.emulator only with the basic blocks executed can ensure the new emulator is compact and efficient in terms of execution time. Basically, the markup of basic blocks starts from the first emulator instruction of a critical clause\footnote{A critical clause is that which counter reached a threshold.} and continues until the last statement of this clause. This process will create a trace that will form a S.emulator. There are two important moments in the system life: when some clause becomes critical, and when the same clause becomes hot. The first indicates the system needs to begin the creation of a S.emulator, and the second indicates that there is a complete trace. In the second moment, the system can finish to reason about its internal structure, compile the new trace and finally install the new S.emulator.

When a trace is initialized, the critical clause and its first basic block executed becomes the head of the trace. After that, each basic block executed is marked to form the S.emulator. The trace consists of the control flow of each emulator instruction, represented by an ordered sequence of basic blocks which begins with a label and finishes with a unconditional branch to the label of the next emulator instruction. It is important to note that a trace does not contain only basic blocks from only one clause. In fact, critical and hot clauses are used to initialize and finalize the construction of a trace. It means a trace does not have knowledge about clauses.

Ideally, an optimal trace should consist solely of basic blocks executed, without conditional statements or useless basic blocks. However, the construction of such trace is not always feasible because it can ensure the efficiency of a program but not the execution of another program. For this reason, each trace, in addition to being efficient in terms of execution time also must ensure the completion of the program and a correct output. These conditions are fulfilled according to the following criteria:

\begin{enumerate}
\item The trace must be appropriate for the data type it was built;
\item The trace must maintain the conditions on the occurrence of exceptions;
\item A dereferenciation must complete successfully; and
\item The indexing instructions must invoke correct clauses.
\end{enumerate}

\paragraph{\bf Handling Data Types}

In practice, all emulator instructions are generic for all data type. Therefore, the system should ensure an efficient trace eliminating conditional statements. However, the Prolog data type can change during runtime. Thus, the system adjusts the trace (more precisely, the S.emulator built previously) to the new data type. Thus, while no change occurs, the trace will have a linear characteristic.

However, if a previously successful conditional statement fails, it is necessary the trace be modified to contain the new basic blocks. This feature, mutability, indicates that during the execution of a program, a trace can be rebuilt to change a previously built S.emulator. The mutability ensures an accurate control flow for all data type.

\paragraph{\bf Handling Exceptions}

In the YAP context, an exception is a control flow that does not belong to a running clause. In practice, exceptions occur during backtracking and garbage collection. Both exceptions are prioritized during the trace construction, which infer that the trace will not ignore the occurrence of such exceptions.

At the implementation level, conditional statements that throw exceptions for garbage collection are not preceded by instructions that evaluate data type and, therefore, constitute a single basic block in the trace. In this case, the basic block represents a conditional statement that, if successful, returns the control flow to the default emulator and invoke the garbage collector. After that, the system triggers the emulators; thus the execution continues from that instruction that throws the exception on the S.emulator.

Moreover, exceptions caused by backtracking are always preceded by conditional statements including instructions for evaluating data type. Unlike exceptions for garbage collection, backtracking are handled by the S.emulator and should not have the conditional statements ignored.

\paragraph{\bf Handling Dereferenciation}

In YAP, the dereferenciation is implemented as a loop to transverse a chain of pointers built during the unification of two variables. This loop finishes when the value of the term is found or when it is determined that a term has not been unified with a value yet. Before this loop is executed, a conditional statement to verify if the term is variable is performed, and the dereferenciation occurs only if this statement conditional is true.

Considering the dereferenciation implemented in YAP, this operation is handled by the proposed strategy in the following ways:

\begin{itemize}
\item If the conditional statement is executed before the dereferenciation loop determines that the term is not a variable, the dereferenciation operation is not inserted in the trace.
\item On the other hand, if the term is variable, a basic block containing the dereferenciation operation is inserted in the trace. This block contains two branches. One will be executed only if the dereferenciation process finishes with a variable, and, therefore, the flow jumps to the basic block which handles variables. The other will be executed if this process founds a non-variable value, in this case the target basic block handles a non-variable.
\end{itemize}

\paragraph{\bf Handling Indexing Instructions}

In practice, the indexing instructions are not profiled as the others emulator instructions and are entire inserted in the trace, even if none of its basic blocks are executed. It is important to note that before a clause be effectively executed, the indexing instructions need to be executed first so that the system can identify the correct clause to invoke. As several stop conditions are related to changing some data type, this condition should be detected by the trace. Therefore, the strategy is to use a conservative construction and maintain the indexing instructions in order to detect these cases, even if such cases never occurred during the construction of the trace. Besides, it is necessary to insert an instruction that performs a return to the default emulator in case of invoked clause has not been inserted into the trace.

In YAP, the indexing instructions are generated on demand \cite{YAPJIT}; thus it is difficult to determine all the clauses that will be invoked after them, which further reinforces the need to use a conservative strategy.

\subsection{Handling the Mutability}

The system recompiles a trace when the execution flow is modified. Normally, every trace has at least an elementary basic block that contains several branches. In these basic blocks is inserted an additional target to allow the control flow returns to the default emulator when it is identified the execution of some basic block taht was not inserted in the trace. Besides, the system sets the two new registers, namely: K and BADDR. The former indicates whether  the return occurred by a elementary basic block or a garbage collection exception. The latter indicates the address from where the trace needs to be rebuilt.

In other words, when a return to the default emulator occurs by an elementary basic block, the system captures the metadata of the current emulator and enables the markup task; thus the new basic blocks executed are inserted in this trace. When the default emulator reaches the head of the trace, this is recompiled, installed and invoked instead of the previous one. The new trace replaces the previous one. It reduces the space cost and avoids the maintenance of a garbage collector to the S.emulator area.

It is necessary to ensure the return to the default emulator, in the case of changing in the behavior of the program. In a situation without mutability, the earlier built S.emulators can become invalid for the new behavior, causing a lot of return to the default emulator, hurting the performance. With mutability, it is possible to change a previously built trace and minimize the overhead of changing the current emulator.

\subsection{Putting It all Together}

\begin{figure}[!h]
  \centering
  \includegraphics[width=.92\textwidth]{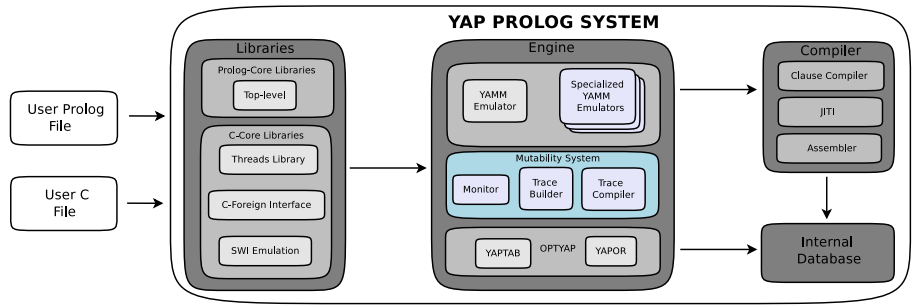}
  \caption{The YAP's architecture with mutability.}
  \label{fig.architecture1}
\end{figure}

The original YAP's architecture comprises four components, namely: \textsc{Libraries}, \textsc{Engine}, \textsc{Compiler} and \textsc{Internal Database}. The \textsc{Libraries} are collections of high and low level libraries responsible for initializing the \textsc{Engine}, supporting threads, native predicates and SWI emulation. The \textsc{Engine} is in charge of executing the Prolog program. The \textsc{Compiler} translates the Prolog code into YAAM instructions. Finally, the \textsc{Internal database} is the database of clauses. A detailed description about the original YAP's architecture can be found in the work of Santos Costa et al. \cite{YAPPROLOG}. In order to provide mutability, the \textsc{Engine} has a new component, called \textsc{Mutability System} that is responsible for creating and handling the S.emulators. Figure \ref{fig.architecture1} provides a general overview of YAP's architecture with mutability.

When a clause becomes critical, the system enables the \textsc{Monitor} that reasons about the internal system structure and marks the basic blocks that were executed. After that, when that clause becomes hot, the system captures the basic blocks, builds a control flow graph, compiles it and finally installs the compiled code (the new S.emulator). Thus, when the engine verifies that the current clause is a head of a S.emulator, it triggers from the default emulator to the correct S.emulator.

The \textsc{Trace Compiler} is a dynamic compiler invoked at runtime which generates a S.emulator from the control flow graph constructed by the \textsc {Trace Builder}. The system uses the
\textsc{LLVM} (Low Level Virtual Machine\footnote{http://www.llvm.org}) \cite{LLVM} to implement the dynamic compiler. The attractiveness of this framework is the fact it generating native code on memory, besides providing a library to tune the process of generating native code.

\section{The Results}
\label{sec.results}

\paragraph{\bf The Experimental Setup}

The experiments in this paper are conducted using {\tt YAP} 
version 6.3 \footnote{http://www.dcc.fc.up.pt/~vsc/Yap} and they are carried
out on an Intel x86\_64 based machine, supporting a 
Intel(R) Xeon(R) CPU E5504 processor running at 2.00GHz, and 24GB of RAM.
The operating system on the machine was Ubuntu, running kernel 3.11.0.
The Table \ref{tab.progs} describes the programs used in the experiments.

\begin{table}[!htb]
  \centering
  \caption{The programs used in the experiments.}
  \label{tab.progs}
  \begin{tabular}{cccccc}
       \hline\hline
      \textbf{Program} & \textbf{Preds.}  & \textbf{Input Size} & \textbf{Program} & \textbf{Preds.} & \textbf{Input Size}\\
      \hline\hline
      append & 1 & 63000000 & hanoi & 1 & 24\\
      nreverse & 2 & 52000 & quicksort & 1 & 52000\\ 
      tak & 1 & 57, 21, 36 & binary trees & 6 & 18 \\
      fannkuch & 11 & 11 & fasta & 13 & 8000000 \\ 
      mandelbrot & 5 & 2400 & n-body & 9 & 3000000\\
      nsieve & 9 & 5 & nsieve bits & 9 & 12\\
      partial sum & 3 & 30000000 & pidigits & 7 & 20000000 \\
      recursive & 4 & 11 & spectral norm & 13 & 800\\
      \hline\hline
    \end{tabular}
\end{table}

The validation of the results is based on the average of 
ten executions. Besides, in the experiments, the
machine workload was minimum as possible, 
in other words, every instance was executed sequential, and
the machine did not have external interference.

The improvement is calculated as follows.
$$Speedup = old\_runtime / new\_runtime$$
$$Improvement = (Speedup-1)*100$$

\paragraph{\bf The Performance}

\begin{figure}[ht]
  \centering
  \includegraphics[width=370pt, height=120pt]{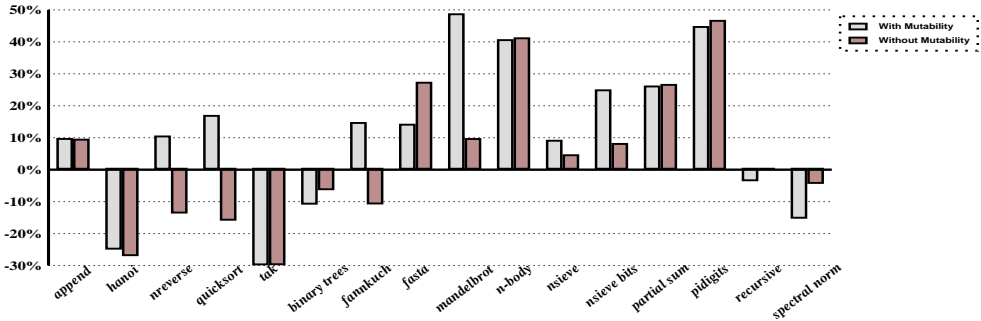} 
  \caption{The improvement of the proposed strategy.}
  \label{fig.stdmem}
\end{figure}

In general, the performance improvement is proportional to the portion of time running on S.emulator. Mutability provided better results with performance improvement up to 10.99\% considering all programs, and 23.57\% considering only program with performance improvement. The Figure \ref{fig.stdmem} shows the performance improvement of the proposed strategy.

The results indicates the proposed strategy achieves better results to programs that contains several predicates. kernels are small programs that have few predicates, besides short runtime. These characteristics impact the performance negatively. In some cases, the performance loss can be addressed by mutability (\texttt{nreverse}, and \texttt{quicksort}), but in general kernels can show slowdown. 

The programs with several predicates obtained better performance, using mutability and without this feature. Only on three programs (\texttt{binary trees}, \texttt{recursive}, and \texttt{spectral norm}), the proposed strategy loses performance.

The results indicate that, in some cases, the mutability can degrade the performance (\texttt{fasta}, \texttt{pidigits}, \texttt{n-body}). In these cases, the return to the default emulator (due to the S.emulator throws an exception) and the decision of rebuilding the S.emulator did not bring the desired effect. It indicates it is difficult to predict the future based on past. However, in general the best choice is to use mutability.

Such results are not only consequences of constructing S.emulators on the fly, but the rules applied on control flow graph construction, which emphasizes the elimination of conditional statements. Building S.emulators without such statements benefits prefetching and branch predictions, and increase the scope of code transformations on a global level. Therefore, these benefits are the reason for achieving results.

\begin{figure}[ht]
  \centering
  \includegraphics[width=370pt, height=120pt]{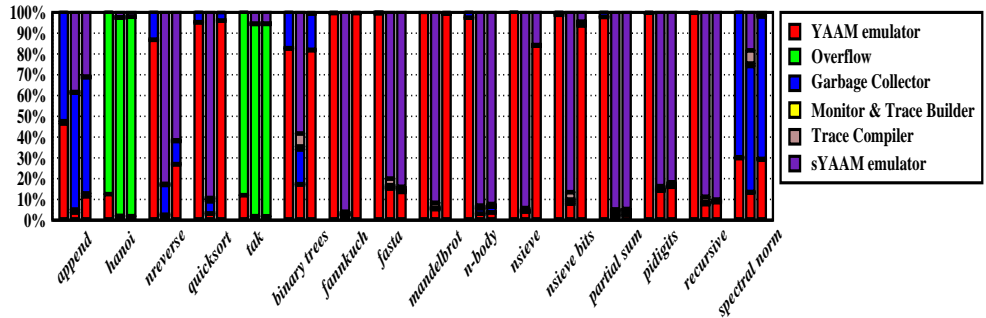} 
  \caption{The breakdown of mutability handling.}
  \label{fig.stdmem_percent}
\end{figure}

To understand the sources of performance loss, it is necessary to evaluate the mutability system in details. So that,
the evaluation described below emphasizes the rate of time spent by the mutability system. In the Figure \ref{fig.stdmem_percent} each bar is composed by several components, namely: default emulator, overflow, garbage collector, monitor e trace builder, trace compiler, and S.emulator. Besides, for each program is shown three bars. The first represents the default emulator, the second represents the proposed system using mutability, and the last the proposed system without mutability.

The cost of monitoring and building traces, as well as, the cost of compilation are minimal and do not degrade the performance. Without mutability, monitoring and building ranged from 0.002\% (\texttt{mandelbrot}) to 0.12\% (\texttt{recursive}) of elapsed time and compiling ranged from just 0.084\% (\texttt{tak}) to 1.496\% (\texttt{pidigits}). These results show that invoking the \textsc{Monitor} only on frequent regions of the program is crucial to achievef performance. Now, with mutability, the cost ranged from 0.005\% (\texttt{hannoi}) to 1.914\% (\texttt{binary trees}) to monitor and build and ranged from 0.073\% (\texttt{hanoi}) to 1.914\% (\texttt{spectral norm}) to compile. This increase is evident by the need to keep the mutability system's modules active for long runtime, but it was not enough to degrade the performance due to the high runtime in S.emulator.
The results, from Figure \ref{fig.stdmem_percent}, indicate that garbage collection, overflow, and the time spend on YAAM emulator are sources of performance loss.

The results indicate it is possible to improve performance even in cases in which the native code is executed for a short time. Such cases include \texttt{mandelbrot} (gain of 9.59\% with S.emulator active at 0.01\% of the time), \texttt{nsieve} (gain of 4.51\% with S.emulator active at 18.18\% of the time), and \texttt{nsieve bits} (gain of 8.06\% with S.emulator active at 4.82\% of the time), but even so, the performance for these was lower than that achieved by mutability. 

\paragraph{\bf Discussion}

The results indicate that the performance gain is not only dependent on the implementation of useful traces, but it can be achieved if useless traces are avoided especially the traces with a high level of useless (\texttt{nreverse} and \texttt{quicksort}, and \texttt{fannkuch}). In this sense, the mutability was very important because, in addition to achieving better performance, in general it minimizes the overhead of useless traces.

Another important point to consider is that, with mutability, the construction and compilation time of traces did not impact the performance of programs that had been improved the performance without mutability, showing that the cost of rebuilding a S.emulator is minimal. The only exception to this rule is the program \texttt{Hanoi}, but for this, the case refers to the exceptions handled only on the default emulator. Additionally, the construction of optimal traces (only) is no guarantee of performance for programs with a high concentration on exception handling.

Finally, based on the results, an important future work is to modify the system so that it addresses the tasks of manipulating the memory areas also on S.emulator in order to ensure efficient implementation of all programs since all useful traces are built. Besides, the task of avoiding useless traces indicates be a good strategy in some cases, but it can be enhanced by the use of global analysis, for example, in order to detect the best starting clause of a trace that is the beginning of an invocation chain.

\section{Concluding Remarks}
\label{sec.conclusion}

This paper proposed the use of introspection in order to create a specialized emulator on the fly. The proposed strategy monitors the emulator execution and then generate a control flow graph in memory only formed by basic blocks executed, which are then compiled and executed with the highest priority.

The proposed strategy did not outperform the default emulator for all programs, due to the overhead of returning to default emulator on the presence of exceptions. Currently, a future work will investigate strategies to minimize this overhead.


\end{document}